\documentclass[twocolumn,aps]{revtex4-2}
\usepackage{amsmath}
\usepackage{amssymb}
\usepackage{graphicx}
\usepackage{dcolumn}
\usepackage{bm}
\usepackage{xcolor}
\usepackage{epstopdf}

\begin{document}

\preprint{Phys.Rev.Lett.}

\title{Diffusion of photo-excited holes in viscous electron fluid}
\author{Yu. A. Pusep$^{1}$, M. D. Teodoro$^{2}$, V. Laurindo Jr.$^{2}$, E. R. C. de
Oliveira$^{2}$, G. M. Gusev$^{3}$, and A. K. Bakarov$^{4}$}
\date{22/02/2022}
\affiliation{$^{1}$S\~{a}o Carlos Institute of Physics, University of S\~{a}o Paulo, PO Box
369,13560-970 S\~{a}o Carlos, SP, Brazil}
\affiliation{$^{2}$Departamento de F\'{\i}sica, Universidade Federal de S\~{a}o Carlos,
13565-905, S\~{a}o Carlos, S\~{a}o Paulo, Brazil}
\affiliation{$^{3}$Institute of Physics, University of S\~{a}o Paulo, 135960-170 S\~{a}o
Paulo, SP, Brazil}
\affiliation{$^{4}$Institute of Semiconductor Physics, 630090 Novosibirsk, Russia}
\email{pusep@ifsc.usp.br}
\date{22/02/2022}

\begin{abstract}
The diffusion of photo-generated holes is studied in a high-mobility
mesoscopic GaAs\ channel where electrons exhibit hydrodynamic properties. It
is shown that the injection of holes into such an electron system leads to the
formation of a hydrodynamic three-component mixture consisted of electrons and
photo-generated heavy and light holes. The obtained results are analyzed
within the framework of ambipolar diffusion, which reveals characteristics of
a viscous flow. Both hole types exhibit similar hydrodynamic characteristics.
In such a way the diffusion lengths, ambipolar diffusion coefficient and the
effective viscosity of the electron-hole system are determined.

\end{abstract}

\maketitle

Specific characteristics of electron transport may strongly influence
performance of micro-electronic devices. In a dense electron system where the
electron-electron collisions dominate over the collisions of electrons with
disorder, the hydrodynamic approach is applied \cite{andreev2011,bandurin2016}%
. These conditions can be fulfilled in high mobility, correlated
two-dimensional electron gas which leads to a variety of new phenomena
associated with the hydrodynamic character of the electron gas, such as
formation of density waves, shock waves, turbulence, solitons etc. Many
related specific effects were predicted and observed: the Gurzhi effect
\cite{gurzhi1963}, choking of the electron fluid
\cite{dyakonov1993,dyakonov1995}, Hagen-Poiseuille charge flow
\cite{molenkamp1994,molenkamp1995}, hydrodynamic pumping effect
\cite{govorov2004} and Hall viscosity \cite{gusev2018PRB98}. Particularly
interesting is the system composed of electron and hole hydrodynamic
components which interact through viscous friction. Hydrodynamic properties of
such a system were considered in \cite{ryzhii2012} where collective
excitations and mutual electron-hole drag were studied. As for the drag
effect, such effects caused by Coulomb interaction have been extensively
demonstrated and investigated in a variety of systems including GaAs quantum
well (QW) \cite{shah1986,shah1988}, double-QW electronic systems
\cite{eisenstein2004} and double-layer graphene \cite{geim2012}. However, to
the best of our knowledge, so far no experiments have been performed in
hydrodynamic electron-hole systems where drag effects are expected to be even
stronger due to a viscous friction between the electron and hole components.

We address our investigation to the diffusion processes which take place in
the hydrodynamic regime in a high-mobility mesoscopic GaAs channel. If the
scattering on the channel edges is diffusive and the mean free path for
electron-electron collisions, $l_{ee} < W$, where W is the channel width, the electron transport should take a form of
the hydrodynamic Poiseuille flow controlled by the electron shear viscosity
$\nu\sim$ v$_{F}$l$_{ee}$, where v$_{F}$ is the Fermi velocity. In particular,
we report on a photocurrent (PC) study of diffusion of the photo-generated
holes within a viscous electron fluid. The question to be answered is: whether
optically injected holes will transform initially hydrodynamic electron system
to a conventional Ohmic system, or a two-component electron-hole hydrodynamic
system will arise.

A single 14 nm thick GaAs/AlGaAs QW was grown on (100)-oriented GaAs substrate
by a molecular beam epitaxy. The sheet electron density and the mobility
measured at the temperature of 1.4 $K$ were 4.8$\cdot$10$^{11}$ cm$^{-2}$ and
1.0$\cdot$10$^{6}$ cm$^{2}$/Vs, respectively. In this structure the viscous
electron transport was demonstrated in
Refs.\cite{gusev2018AIP,gusev2018PRB97,gusev2018PRB98}.

Scanning PC microscopy experiments were performed on a multi-terminal Hall bar
structure with the 5 $\mu$m width and 100 $\mu$m length of the active at the
temperature 3.7 K using a helium closed cycle cryostat equipped with a
superconducting magnet (Attocube/Attodry1000). An electrically connected
sample was placed on top of a x-y-z nanopositioner stack (Attocube), which
allows for precise positioning of the laser beam focused by an aspheric
objective (NA=0.64) along the channel. The 532 nm illumination with the pump
power about 0.5 mW from the laser (Cobolt/08) was focused onto the sample. The
laser spot size is about 2 $\mu$m. The PC\ measurements were carried out by a
source meter Keithley 2400. Time-resolved photoluminescence\ (PL) measurements
were performed with the optical excitation achieved by a Pico Quant/LDH Series
diode laser emitting 80 MHz pulses at 730 nm operated at an average pump power
of 5 $\mu$W corresponding to a peak power of about 0.8 mW, which was chosen to
generate the same number of electron-hole pairs as in the case of the
continuum laser. In this case emission was dispersed by a 75 cm Andor/Shamrock
spectrometer and the PL transients were detected by a PicoQuant Hybrid PMT
detector triggered triggered with a time correlated single photon
PicoQuant/PicoHarp 300 counting system.

During the measurements the laser spot was scanned along the channel. The PC
was measured using the potential probes of the Hall bar. Such configuration is
usually employed to determine the diffusion length of the optically excited
minority carriers: diffusion results in the PC value increasing with the
decreasing distance between the probes and the laser spot. The solution of the
one dimensional diffusion equation has a Gaussian shape with the exponent
defined by the diffusion length. In such a case the diffusion length is
obtained by fitting the calculated and measured PCs as a function of the distance.%

\begin{figure}[ht!]
\includegraphics[width=9cm]{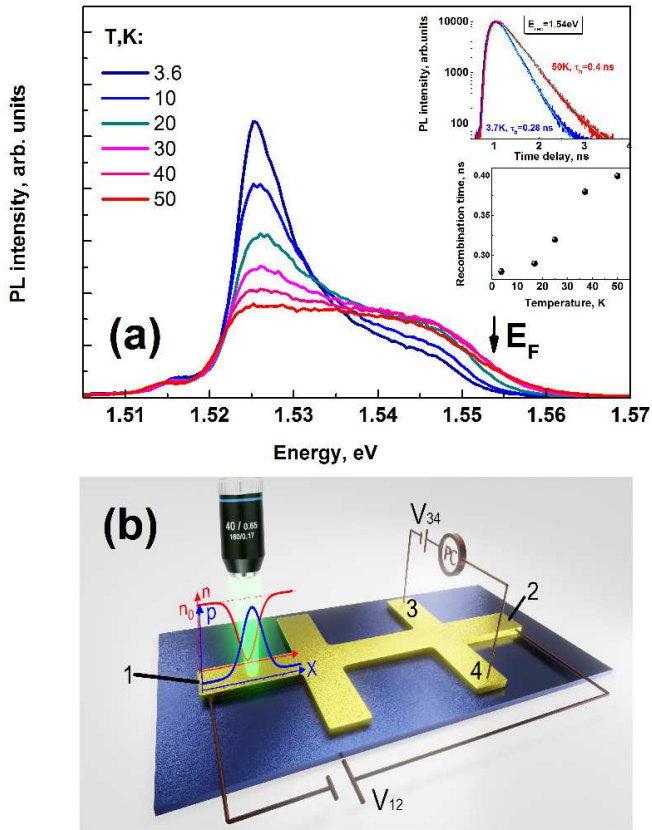}
\caption{\label{fig1}(Color online)(a) PL spectra of the GaAs/AlGaAs QW measured at various
temperatures. Transients measured at the PL energy 1.54 eV at the point close
to the collecting probes at T=3.7K (blue lines) and T=50K (red lines) are
shown in the top inset, where the cyan lines are the best fits using a
mono-exponential decay function; the recombination times obtained by the best
fits are shown in the bottom inset. (b) A sketch of the experimental
configuration demonstrates electron $n$ and hole $p$ densities after
photoexcitation as a function of the position along the channel and the
electronic circuit that was attached to the sample; $n_{0}$ is the background
electron concentration. }
\end{figure}

The optical characterization of the QW structure studied here is shown in
Fig.1. The PL spectra depicted in Fig.1(a) reproduce the joint density of
states with the conduction band states populated below the Fermi level. The
observed PL is determined by the recombination of the conduction band states
below the Fermi level with holes in the valence band. In this case, the
spectral width of the PL should be approximately equal to the Fermi energy.
Indeed, the width of the PL spectra shown in Fig.1(a) is in good agreement
with the Fermi energy of 30 meV obtained in the same sample from
magnetotransport measurements in \cite{gusev2018PRB97}. The results of the PL
time-resolved measurements shown in the panels (a,b) reveal the recombination
time increasing with temperature, what is expected in QWs with a degenerate
electron system due to the phonon-assisted Auger recombination
\cite{polkovnikov1998}. No significant variation of the recombination \ time
along the channel is detected. The observed recombination time is attributed
to the optical transitions between the conduction band and heavy hole valence
band confined states.

A sketch of the experimental configuration is presented in Fig.1(b). The
electron and photogenerated hole densities after photoexcitation are shown as
a function of the position along the mesoscopic channel. In the GaAs/AlGaAs
heterostructure studied here, electron-hole pairs photogenerated in AlGaAs
barriers are separated by a potential at the QW heterointerfaces: holes pass
into the QW, while electrons nonradiatively recombine in the barriers. The
recombination of holes injected in this way with background electrons reduces
their concentration in the channel and, consequently, results in a minimum of
the photocurrent measured between probes 3 and 4. Such a mechanism was found
in Refs.\cite{chaves1986,richards1990,hayne1994}.

The observed decrease in PC with a decrease in the distance between the laser
spot and the collecting probes, shown in Fig.2(a), manifests to a dominant
role of recombination between holes and electrons injected into the channel.
Such a process is expected in high-mobility QWs where the recombination time
$\tau$ is shorter than the electron transport relaxation time. Thus, the
system under study consists of the background electrons and the
photo-generated holes, in which diffusion proceeds in an ambipolar form.

The Hall bar sample studied here shown in Fig.2(b) has eight potential probes.
All the potential probes revealed identical PC responses. The PC measured as a
function of the laser spot position along the channel using the probes 3,4 and
5,6 are depicted in Fig.2(a). In the following, the data obtained with the
collecting probes 3,4 are demonstrated.

Worth mentioning that the current measured between the potential probes shown
in Fig.2(a) consists of the current due to the voltage applied to the probes
and the photocurrent itself. However, throughout the article we will call this
current photocurrent, because it directly demonstrates the effect of
electron-hole recombination.

To prove the reduction in the number of electrons observed under laser
illumination, we measured the Hall voltage across the collecting contacts
without and with laser excitation, when the laser was focused on the channel
area between the contacts. The obtained data shown in Fig.2(d) indeed confirm
a twofold decrease in the electron concentration under illumination.%

\begin{figure}[ht!]
\includegraphics[width=9cm]{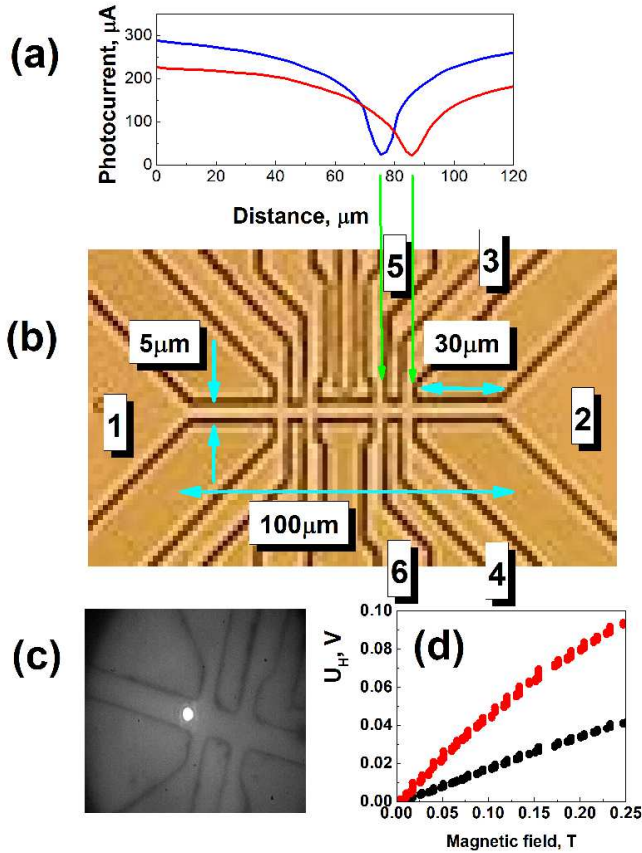}
\caption{\label{fig.2}(Color online)a) Photocurrent measured at T=3.7K with collecting
probes 34 (red line) and 56 (blue line), (b) microscope image of the sample
with the characteristic dimensions, (c) the sample image with the laser spot
focused to the channel and (d) Hall voltages U$_{H}$ measured on collecting
probes 3 and 4, at a current I$_{12}$ = 0.13 mA between the probes 1 and 2, in
the dark (black circles) and at an average laser pump power of 5 $\mu$W (red
circles)}
\end{figure}

The results shown in Fig.2(a,d) imply the concentration of holes injected to
the QW, comparable to the electron background concentration. In such a case
the diffusion takes a form of an ambipolar electron-hole diffusion.

As stated above, the specific scattering conditions establish the viscous
electron flow in the sample studied here. At the same time it is unclear
whether such conditions apply to photo-generated holes. In order to define a
character of the photo-generated hole diffusion the diffusion profiles were
measured at different temperatures.

In the hydrodynamic approach the electron-electron, hole-hole, and
electron-hole scattering dominates over the scattering of electrons and holes
with disorder. In such a case diffusion reveals characteristics of viscous
flow: the diffusion length should increase with the increasing temperature. In
the case considered here, an increase in the diffusion length of the
photo-generated holes with temperature will show that they exhibit
hydrodynamic properties.

The PC measured at different temperatures shown in Fig.3 is determined by the
doping electron concentration, which decreases due to recombination with holes
when they arrive the collecting contacts. The net electron concentration that
contributes to the PC can be expressed as $n(x)=n_{0}-n_{h}(x)$, where $n_{0}$
is the doping background electron concentration in the absence of
photo-excitation and $n_{h}(x)$ is the concentration of photo-generated holes,
which is represented by the Gaussian profile. As shown in Fig.3(a), the PC
calculated in this way as a function of the distance between the laser spot
and the collecting contacts does not fit the experimental diffusion profile. A
considerable deviation is found at long distances, which is a signature of an
additional diffusion flow. A possible cause of this additional flow can be
presence of light holes generated by the light.%

\begin{figure}[ht!]
\includegraphics[width=9cm]{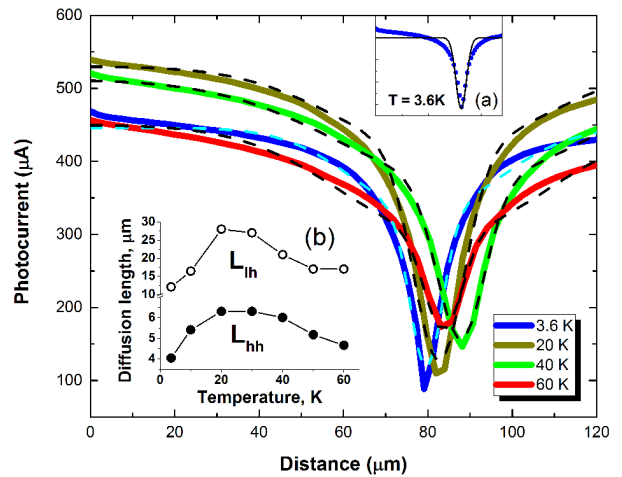}
\caption{\label{fig.3}(Color online)Photocurrent measured at different temperatures at
U$_{34}$=2 V as a function of the distance along the channel (solid lines).
The dashed lines were calculated according to Eq.(1) with the contributions of
both heavy and light holes. The panel (a) shows the fit of the photo-current
measured at T=3.6K by Eq.(1) with the contribution of only heavy holes. The
panel (b) shows the diffusion lengths of heavy and light holes as a function
of the temperature}
\end{figure}

Following the approach presented above, the net electron concentration can be
calculated according to the formula:%

\begin{equation}
n(x)=n_{0}-n_{hh}\exp(-\frac{x^{2}}{4L_{hh}^{2}})-n_{lh}\exp(-\frac{x^{2}%
}{4L_{lh}^{2}}) \label{1}%
\end{equation}
where the second and third terms are associated with the heavy and light holes
arriving at the collecting probes, $n_{hh(lh)}$ is the heavy hole (light hole)
concentrations, while $L_{hh}$ and $L_{lh}$ are the diffusion lengths of the
heavy and light holes, respectively. Then the PC as a function of distance can
be calculated as j$_{PC}$(x) $=$ en(x)v$_{F}$. A good accordance between the
experimental data and the PC calculated by Eq.(1) manifests to the
simultaneous diffusion of the heavy and light holes. The diffusion lengths
obtained by the best fits are shown in the panel (b) as a function of
temperature. The shorter diffusion length is associated with the heavy hole
flow, while the longer one is due to the light holes. Up to 30K both diffusion
lengths increase with increasing temperature, while higher temperatures result
in decreasing of the diffusion lengths. The diffusion length increasing with
temperature manifests itself in a viscous flow. For temperatures higher than
about 35 K, strong polar LO-phonon scattering determines the momentum
relaxation time in GaAs QWs. In this case phonon scattering dominates over the
electron-electron collisions and diffusion takes a common form.

Thus, the presented results demonstrate formation of the electron-hole plasma
composed by the electrons and injected holes in the studied here mesoscopic
GaAs channel. At low temperatures diffusion of this plasma reveals viscous character.

The PC measured as a function of the distance with different voltages U$_{34}$
applied to the probes 3 and 4 is shown in Fig.4(a). The best fits of the
experimental PC diffusion profiles using Eq.(1) are shown in Fig.4(a). The
obtained diffusion lengths $L_{hh}$ and $L_{lh}$ are found slightly decreasing
with the increasing U$_{34}$ voltages.%

\begin{figure}[ht!]
\includegraphics[width=9cm]{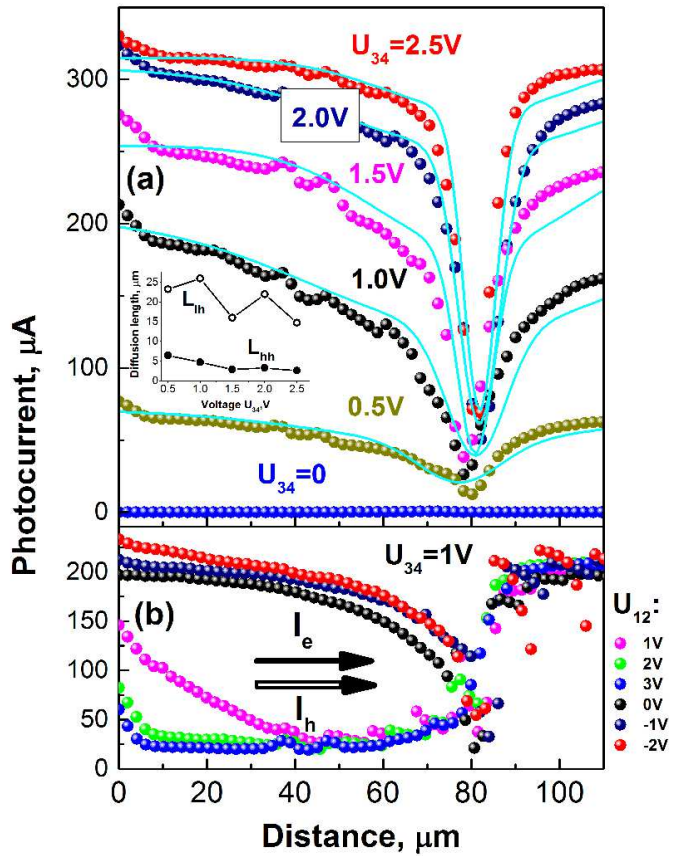}
\caption{\label{fig.4}(Color online) Photocurrent measured at T = 3.7 K as a function of
distance along the channel with different U$_{34}$ (a) and U$_{12}$ (b)
voltages. The solid lines in the panel (a) are calculated according to Eq.(1),
while the diffusion length of heavy and light holes are depicted in the inset.
The directions of the electron drift I$_{e}$ and the hole flow I$_{h}$ along
the channel are shown corresponding to the positive U$_{12}$ voltage.}
\end{figure}

In the following the effect of the electron-hole drag is investigated using an
external electric field applied parallel to the diffusion flow. The
application of the longitudinal voltage U$_{12}$ makes it possible to control
the number of photo-generated carriers arriving at the collecting probes.
Diffusion profiles obtained with different U$_{12}$ values when negative and
positive potentials are applied to contacts 1 and 2, respectively, and with
fixed U$_{34}$ =1V are shown in Fig.4(b). In this case an apparent asymmetry
of the diffusion profile is found. Such a change of the diffusion profile
reveals a drift of electrons in the external electric field which drag
photo-generated holes. From the left side of the collecting probes, the flow
of electrons carries holes to the collecting contacts, facilitating their
diffusion. In this case, holes located farther from the collecting probes
contribute to PC. While on the right side, electrons carry holes away from the
contacts. Accordingly, drift of the injected photo holes against the electric
field is observed due to their entrainment by the drift electron current.

It should be noted that the drift of the minority electrons against the
electric field caused by negative absolute mobility was observed in p-doped
GaAs QWs in \cite{shah1986,shah1988}. Moreover, it was argued that this should
result in the negative photoconductivity. Such an effect indeed may reduce the
photoconductivity in the case when the momentum relaxation time is much
shorter than the electron-hole recombination time. However, in the samples
studied here the high electron mobility implies the momentum relaxation time
longer than or equal to the recombination time. In these conditions the
recombination process dominates over the transport which suggests the
recombination as a principal reason for the observed decrease in PC. In
addition, according to \cite{shah1988}, the negative photoconductivity is not
expected at high excitation, when the density of photo-generated carriers
approaches the density of the majority carriers.

It is interesting to note that in this case the PC diffusion profile reveals
asymmetric shape. This performance of the observed electron-hole drag is very
different from that found in the diffusion electron-hole system
\cite{shah1986,shah1988}, where Gaussian-like diffusion profiles were
observed. The observed asymmetry of the diffusion shape implies a very strong
drag effect, which is likely caused by additional viscous friction between
electrons and holes.

It is worth noting that the observed drag effect can explain the diffusion
length decreasing with the increasing voltage applied to the collecting
probes, shown in the inset to Fig.4(a). Namely, the current flowing through
the collecting contacts entrain the holes arriving the contacts resulting in
easier diffusion.

With the purpose to describe the observed diffusion the following analysis is
performed. The diffusion length of the viscous particles can be calculated
following the formalism presented in Ref.\cite{achterberg2016}. The fluid
viscosity leads to shear flow along the x-direction with a velocity $v_{x}(y)$
which varies in the y-direction. The diffusion transfers momentum. In a
one-dimensional shear flow the momentum density only has an x-component:%

\begin{equation}
M_{x}=\rho v_{x}(y) \label{2}%
\end{equation}
where $\rho$ is the fluid density. In the considered case the momentum flux is
given by the expression:%

\begin{equation}
T_{xy}=D(\frac{\partial M_{x}}{\partial y})=\rho D(\frac{\partial v_{x}%
}{\partial y}) \label{3}%
\end{equation}
where $D$ - is the diffusion coefficient. On the other hand, by a definition a
viscous flux $T_{xy}=\eta(\partial v_{x}/\partial y)$ with the dynamic
viscosity coefficient $\eta$. As a result, $\eta=\rho D$. The kinematic
(shear) viscosity of a fluid is defined as $\nu\equiv\eta/\rho$. Thus, in the
case of a one-dimensional diffusion the kinematic viscosity $\nu$\ and the
diffusion coefficient of the involved particles are the same quantity
\cite{achterberg}. Finally, the diffusion length of the viscous fluid is
$L_{D}$ =$\sqrt{D\tau}$ where $\tau$ is the particle lifetime (in the case
studied here it is the recombination time).

In the case of the ambipolar diffusion studied here the diffusion length is
attributed to the mutual diffusion of electrons and holes. Using the diffusion
length $L_{hh}=5$ $\mu$m, together with the heavy hole recombination time
$\tau=$ 2.8$\cdot$10$^{-10}$ s yield the ambipolar diffusion coefficient and
consequently, the effective viscosity coefficient $D_{a}$ = $\nu_{a}$ = 0.09
m$^{2}$/s. The obtained effective viscosity coefficient is found 3 times
smaller than that of electrons $\nu_{e}$ $\approx$ 0.3 m$^{2}$/s, determined
in the same structure by nonlocal electrical measurements
\cite{gusev2018PRB97}.

In summary, the diffusion of photo-generated holes was studied by means of a
scanning PC microscopy carried out on a mesoscopic channel formed in a
high-mobility GaAs QW where the electrons reveal hydrodynamic properties. As a
result, the diffusion of the photo-generated holes was investigated within the
viscous fluid of the electrons. It was shown that the hole system consists of
heavy and light holes. Diffusion lengths of both hole types were obtained.
Depending on the temperature, two diffusion regimes were observed:
hydrodynamic regime at temperatures below 30K and common diffusion at higher
temperatures. Both types of holes exhibited similar hydrodynamic properties.
The drift of photo-generated holes against external electric field was found
due to the hole drag by viscous electrons. In this case the corresponding
diffusion profiles were found very different from those in the usual diffusion
electron-hole system. The obtained results were analyzed within the framework
of ambipolar diffusion. In such a way the ambipolar diffusion coefficient and
the effective viscosity of the electron-hole system were determined. The
presented results differ from the hydrodynamic effects observed so far in
viscous electron systems, since in the reported case the diffusion of holes
occurs within a mixture consisting of the hydrodynamic electrons and the
injected photo holes. Thus, the obtained results manifest to the formation of
a three-component hydrodynamic system formed by background electrons and
injected heavy and light holes.

Acknowledgments: Financial supports from the Brazilian agencies FAPESP (Grant
2015/16191-5), CNPq (Grant 305837/2015-0), CAPES (Grant PNPD
88887.336083/2019-00) are gratefully acknowledged.

\end{document}